\documentclass[conference]{IEEEtran}
\IEEEoverridecommandlockouts
\usepackage{hyperref}
\usepackage{amsmath,amssymb,amsfonts}
\usepackage{cite}
\usepackage{dutchcal}
\usepackage[T1]{fontenc}
\usepackage{subfigure}
\usepackage{textcomp}
\usepackage{xcolor}
\usepackage{hyperref}
\usepackage{booktabs}
\usepackage{footnote}
\usepackage{threeparttable}
\usepackage{multirow}
\usepackage{algorithm}
\usepackage{algorithmic}
\usepackage{graphicx}
\usepackage{bbding}
\usepackage{makecell}
\usepackage{stfloats}
\usepackage{url}
\usepackage{verbatim}
\usepackage{microtype}
\usepackage[mathscr]{eucal}
\usepackage[capitalize,noabbrev]{cleveref}
\makeatletter
\newcommand{\linebreakand}{%
  \end{@IEEEauthorhalign}
  \hfill\mbox{}\par
  \mbox{}\hfill\begin{@IEEEauthorhalign}
}

\newcommand{\Rmnum}[1]{\expandafter\@slowromancap\romannumeral #1@}
\makeatother
\def\BibTeX{{\rm B\kern-.05em{\sc i\kern-.025em b}\kern-.08em T\kern-.1667em\lower.7ex\hbox{E}\kern-.125emX}}
\begin{document}

\title{Multi-Modal Representation Learning for Molecular\\ Property Prediction: Sequence, Graph, Geometry
\thanks{This paper is submitted to 2023 IEEE International Conference on Bioinformatics and Biomedicine (BIBM).}
}

\author{
\IEEEauthorblockN{1\textsuperscript{st} Zeyu Wang}
\IEEEauthorblockA{\textit{Institute of Cyberspace Security} \\
\textit{Zhejiang University of Technology}\\
Hangzhou, China \\
vencent\_wang@outlook.com}
\and
\IEEEauthorblockN{2\textsuperscript{nd} Tianyi Jiang}
\IEEEauthorblockA{\textit{Institute of Cyberspace Security} \\
\textit{Zhejiang University of Technology}\\
Hangzhou, China\\
josieyi0319@163.com}
\linebreakand
\IEEEauthorblockN{3\textsuperscript{rd} Jinhuan Wang}
\IEEEauthorblockA{\textit{Institute of Cyberspace Security} \\
\textit{Zhejiang University of Technology}\\
Hangzhou, China\\
jhwang@zjut.edu.cn}
\and
\IEEEauthorblockN{4\textsuperscript{th} Qi Xuan$^*$}
\IEEEauthorblockA{\textit{Institute of Cyberspace Security} \\
\textit{Zhejiang University of Technology}\\
Hangzhou, China\\
xuanqi@zjut.edu.cn}
}

\maketitle
\begin{abstract}
Molecular property prediction refers to the task of labeling molecules with some biochemical properties, playing a pivotal role in the drug discovery and design process. Recently, with the advancement of machine learning, deep learning-based molecular property prediction has emerged as a solution to the resource-intensive nature of traditional methods, garnering significant attention. Among them, molecular representation learning is the key factor for molecular property prediction performance. And there are lots of sequence-based, graph-based, and geometry-based methods that have been proposed. However, the majority of existing studies focus solely on one modality for learning molecular representations, failing to comprehensively capture molecular characteristics and information. In this paper, a novel multi-modal representation learning model, which integrates the sequence, graph, and geometry characteristics, is proposed for molecular property prediction, called SGGRL. Specifically, we design a fusion layer to fusion the representation of different modalities. Furthermore, to ensure consistency across modalities, SGGRL is trained to maximize the similarity of representations for the same molecule while minimizing similarity for different molecules. To verify the effectiveness of SGGRL, seven molecular datasets, and several baselines are used for evaluation and comparison. The experimental results demonstrate that SGGRL consistently outperforms the baselines in most cases. This further underscores the capability of SGGRL to comprehensively capture molecular information. Overall, the proposed SGGRL model showcases its potential to revolutionize molecular property prediction by leveraging multi-modal representation learning to extract diverse and comprehensive molecular insights. Our code is released at \url{https://github.com/Vencent-Won/SGGRL}.
\end{abstract}

\begin{IEEEkeywords}
Sequence, Graph, Geometry, Molecular Representation Learning
\end{IEEEkeywords}

\section{Introduction}

Drug discovery in traditional laboratories is a challenging, expensive, and exceedingly time-consuming endeavor~\cite{schneider2020rethinking}. A candidate compound can be utilized in drug design only after undergoing comprehensive testing of its physical, biological, and chemical properties. Consequently, molecular property prediction stands as a pivotal stage in the drug discovery and design process, involving the determination of biochemical properties of unknown compounds~\cite{li2022deep}. In recent years, with the rapid advancement of artificial intelligence, the utilization of AI technology in drug discovery and design has garnered significant attention due to its potential to streamline the discovery cycle and reduce costs effectively~\cite{walters2020applications}. Within this landscape, deep learning-based molecular property prediction methods have showcased remarkable performance. And there are lots of data-driven molecular representation learning methods, which can be mainly divided into three categories~\cite{fang2022geometry}: sequence-based~\cite{zhang2018seq3seq}, graph-based~\cite{wieder2020compact}, and geometry-based~\cite{fang2022geometry} representation methods.
\begin{figure}
    \centering
    \includegraphics[width=\linewidth]{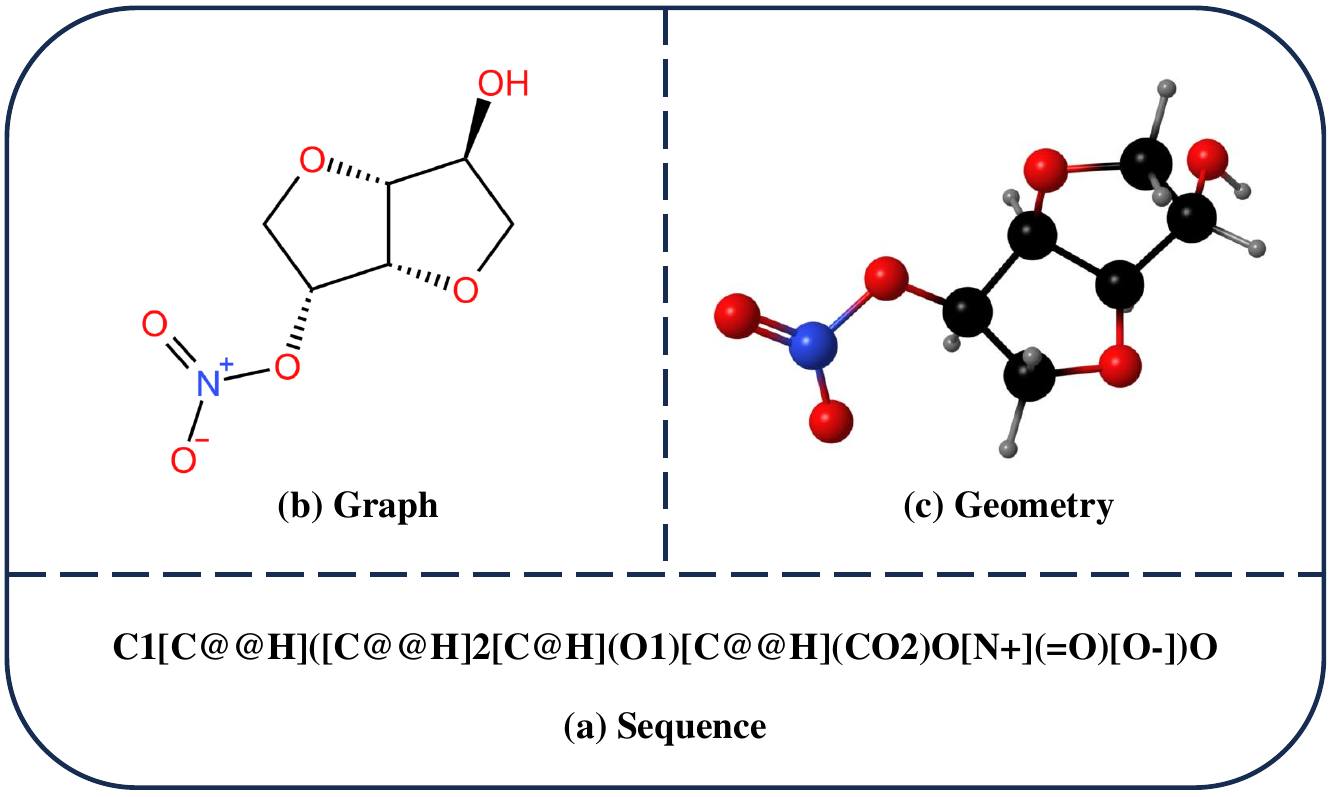}
    \caption{The examples of sequence, graph, and geometry modal of a molecule.}
    \label{fig: modal examplify}
\end{figure}

In sequence-based methods, a molecule can be represented by the Simplified Molecular Input Line Entry System (SMILES) with ASCII strings~\cite{weininger1988smiles}. And sequence-based methods are mainly concerned with nature language processing fields that extract the characteristics of each string to capture broader global information. Examples of such methods include RNNSeq2seq~\cite{xu2017seq2seq}, Smiles-Bert~\cite{wang2019smiles}, Smiles-Transformer~\cite{honda2019smiles}, etc. As illustrated in \cref{fig: modal examplify}~(a), SMILES encoding adheres to principles that atoms, bonds, chemical valences, clearly functional groups composition (such as [N+](=O)[O-], etc.), and chirality information (such as @, /, etc.). Therefore, sequence-based methods can effectively learn molecular representation. Although sequence-based methods can provide the unique encoding for each periodic, bond and hyphen, their representation capability is significantly restricted by the absence of molecular topology information~\cite{ijcai2020p392}.

On the other hand, a molecule can be naturally represented as a graph, where the nodes denote atoms and the edges denote bonds, as illustrated in \cref{fig: modal examplify}~(b). And the molecular graph provides the topology of the molecule such as atom connectivity, the number and size of rings, etc. In contrast to sequence-based methods, graph-based methods exploit the topology of molecules. Graph neural networks methods~\cite{gilmer2017neural, ijcai2020p392, gasteiger2021gemnet}, for instance, extract molecular representation by aggregating neighbor information and have demonstrated their efficacy in tasks like molecule generation~\cite{Shi2020GraphAF} and molecular property prediction~\cite{hao2020asgn}. However, graph-based methods may struggle with graphs that have similar topology but different properties. Moreover, molecular graphs do not contain important information such as chirality, molecular conformation~(as is shown in \cref{fig: modal examplify}~(c)), etc. 

To face these challenges, researchers have turned their attention to developing geometry-based representation methods, such as~\cite{stark20223d,fang2022geometry,wang2023automated}. But, the geometry-based methods pay more attention to geometric-level molecular information. And the conformation is always generated by chemical tools like RDKit, etc. Consequently, there is a need for methods that harness different modalities to achieve a more comprehensive molecular representation. Moreover, certain methods combine two modalities to extract molecular representation. Such as GraSeq~\cite{guo2020graseq} utilizes the sequence and graph modals but neglects the geometric information; GeomGCL~\cite{li2022geomgcl} uses the geometric modal as the augmentation view but neglects chirality, which the sequence modality incorporates. 

\begin{figure*}[!t]
    \centering
    \includegraphics[width=\linewidth]{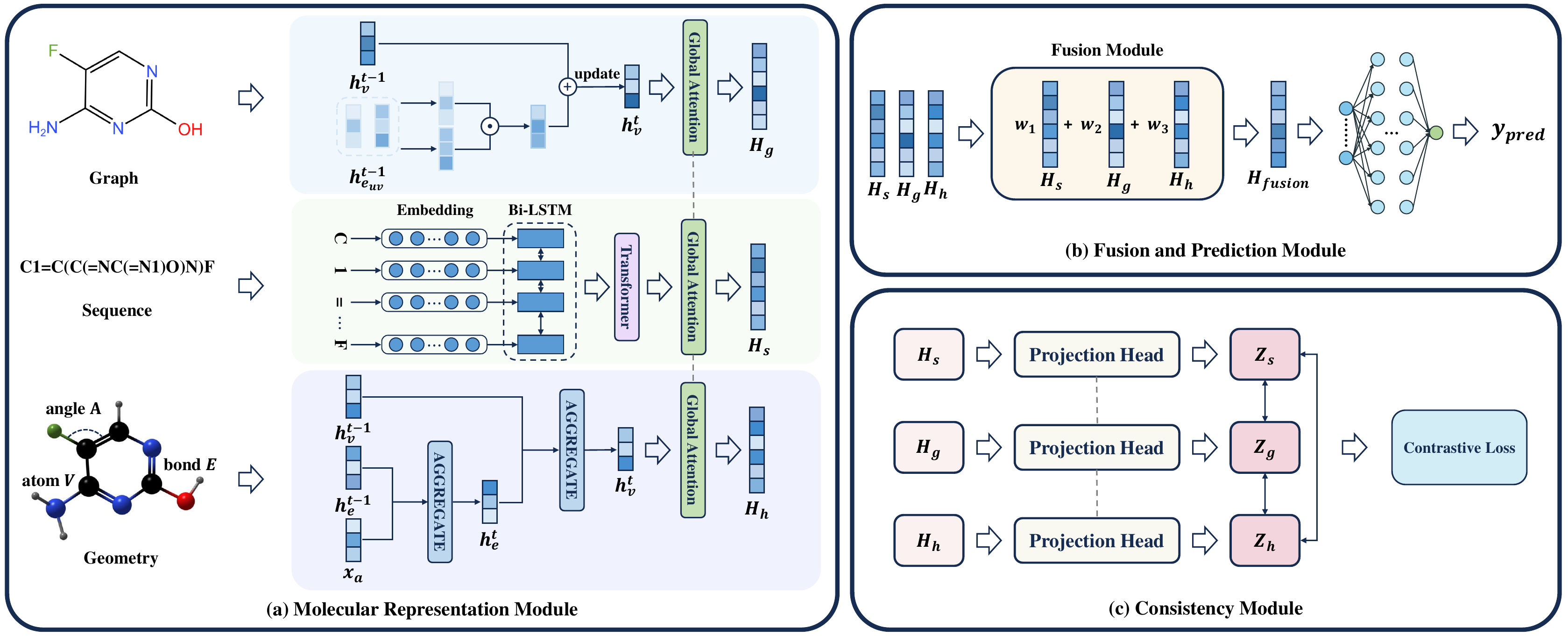}
    \caption{Overview of SGGRL.}
    \label{fig: Overview of SGGRL}
\end{figure*}

To address the above problems, this paper proposes a novel multi-modal molecular representation learning model that integrates the SMILES sequence, molecular graph, and molecular geometry modal characteristics, called SGGRL. Specifically, SGGRL first utilizes a sequenced encoder, graph encoder, and geometric encoder with GlobalAttentionPool layers to generate different modal representations. To ensure modality consistency and reduce information redundancy, SGGRL incorporates a contrastive learning mechanism. This mechanism maximizes the similarity between representations of the same molecule while minimizing the similarity between representations of different molecules. And the representations of the three modalities will be mixed by the fusion layer to acquire comprehensive molecular representations for downstream tasks. The overview of SGGRL is shown in \cref{fig: Overview of SGGRL}. The main contributions of this paper are as follows:
\begin{itemize}
    \item Diverging from prior research, we introduce a novel multi-modal molecular representation learning model that incorporates sequence, graph, and geometry modalities for molecular property prediction.
    \item Considering the sequence is undirect, we introduce the bidirectional LSTM (Bi-LSTM) to substitute the sequence embedding layer and position embedding layer, which can capture the SMILES contextual information. Additionally, we incorporate GlobalAttentionPool and fusion layers into molecular-level representation learning and modality fusion. These enhancements facilitate the model in capturing essential information while reducing redundancy.
    \item Extensive experiments on benchmark molecular property prediction datasets are conducted. Notably, SGGRL significantly outperforms all baselines and achieves $97.9\%$ on Clintox and $96.7\%$ on BBBP.
\end{itemize}

\section{Related Work}
In this section, we review related work on sequence-based, graph-based, geometry-based, and multi-modal molecular representation learning methods. Molecular representation learning is the crucial basis of molecular property prediction. The early works mainly utilized some statistical methods and handcrafted features, such as molecular fingerprint~\cite{cereto2015molecular}, Coulomb Matrix~\cite{rupp2012fast}, etc. However, these methods lack generalizability and scalability. For instance, fingerprint-based methods rely heavily on predefined fingerprint dictionaries. In recent years, data-driven methods that combined with deep learning to extract molecular representation, have received lots of attention, which can be divided into three pipelines: sequence, graph, and geometry~\cite{fang2022geometry}. 

\subsection{Single-Modal Molecular Representation Learning Methods}
Sequence-based methods represent a molecule as a SMILES string and model molecular representation learning as a natural language processing task. For example, Sa-BiLSTM~\cite{zheng2019identifying} introduced the self-attention mechanism to learn important substructure strings (sub-molecules) for extracting molecular representation. Similarly, SMILES-Transformer~\cite{honda2019smiles} proposed a pre-trained method on large-scale molecules using the transformer architecture. However, the lack of structural information limits the performance of sequence-based methods. In contrast, graph-based methods model molecules as molecular graphs and exploit molecular topology. Most graph neural networks are variants of the Message Passing Neural Network~\cite{gilmer2017neural}, utilizing message-passing mechanisms to aggregate neighbor information and capture topological features, such as D-MPNN~\cite{DMPNN2019}, CMPNN~\cite{ijcai2020p392}, etc. Despite their advancements, graph-based methods still face limitations in capturing geometric information, such as molecular conformation. Researchers utilize geometric coordination and angles to model the molecular conformation and extract the molecular representation from it. GEOM~\cite{axelrod2022geom} uses advanced sampling and semi-empirical density functional theory (DFT) to construct molecular structures and predict molecular characteristics. Similarly, GEM~\cite{fang2022geometry} introduces a geometry-based neural network architecture designed to learn molecular geometry knowledge.

\subsection{Multi-Modal Molecular Representation Learning Methods}
Obviously, single-modal methods can not comprehensively describe a molecule. Factors such as chirality, functional groups, scaffold, chemical valency, molecular topology, and conformation information are the impact factors of the molecular properties. Therefore, multi-modal molecular representation is needed. For example, GraSeq~\cite{guo2020graseq} proposes a complementary combination of graph neural networks and uses recurrent neural networks for modeling two types of molecular inputs with the combination of supervised loss and unsupervised loss for training. GET~\cite{mao2021molecular} introduces the Graph Enhanced Transformer, which fuses SMILES representations with atom embeddings learned from Graph Neural Networks to address the retrosynthesis prediction task. In contrast to the sequence-graph combination, GeomGCL~\cite{li2022geomgcl} and GraphMVP~\cite{liu2022pretraining} establish correspondence and consistency between the graph and geometry modalities for pre-training molecular representation models. They leverage conformational information as a complement to the graph modality. However, whether in sequence-graph fusion or graph-geometry fusion, one modality is always lacking to fully represent a molecule. Above all, this paper proposes SGGRL, which distinguishes itself from the aforementioned works by combining sequence, graph, and geometry modalities to learn a more comprehensive representation.
 
\section{Methodology}
As shown in \cref{fig: Overview of SGGRL}, the SGGRL framework is composed of three primary modules: the Molecular Encoder Module, the Fusion and Prediction Module, and the Consistency Module. In order to comprehensively capture molecular characteristics, SGGRL employs the Molecular Encoder Module. This module incorporates a sequence-based encoder, a graph-based encoder, and a geometry-based encoder, enabling the extraction of valuable insights from distinct modalities. To enhance the SMILES encoding and avoid the directional effects of sequence modal, the Bi-LSTM is introduced to replace the embedding layer and position encoding layer. This addition serves to enrich the learning of SMILES contextual information. To ensure compatibility and consistency among different modalities while minimizing redundancy, a fusion layer is strategically integrated, accompanied by the introduction of a consistency learning mechanism. In this section, we will provide a detailed breakdown of the SGGRL framework, elucidating its components and functionalities.

\subsection{Problem Formulation}\label{Problem Formulation}
The purpose of molecular property prediction is to match the corresponding biochemical properties of objective molecules. While molecular representation is needed for downstream property prediction tasks. Therefore, molecular representation learning aims to utilize the representation learning models to acquire accurate molecular representation with molecular encoding information. Formally, one can define the molecular property prediction problem as:
\begin{equation}
    y = \mathcal{F}(\mathbf{Encoder}(X))
    \label{Problem definition}
\end{equation}
where $y$ is the target property, $\mathcal{F}$ is the molecular property prediction predictor, $\mathbf{Encoder}$ is the representation learning model, and $X$ is the molecular encoding input. In this work, the molecular encoding input can be extracted by the SMILES sequence, molecular graph, and molecular geometry.

\subsection{Molecular Encoder Module}\label{Encoder}
The molecular encoder of SGGRL consists of three modal encoders: sequence encoder, graph encoder, geometry encoder, and readout layer. The details of the SGGRL encoder are as follows.

\textbf{Sequence Encoder.} As shown in \cref{fig: Overview of SGGRL}~(a), here, SGGRL introduces a transformer-based molecular encoder. sequence-based molecular property prediction methods utilize the SMILES with nature language models to learn molecular representations. 
First, a molecule can be represented as a SMILES sequence $S=\{s_i|i=1,\cdots,s_T\}, s_i\in D$, where $s_i$ is a token of a SMILES string, $T$ is the length of the SMILES sequence and $D$ is the dictionary of SMILES. Different from other transformer-based encoders, SGGRL utilizes the one-hot encoding to map the SMILES sequence to the feature vectors $X_{seq}=\{x_{s_i}|i=1,\cdots,t\},x_{s_i}\in\mathbb{R}^{|D|}$. Given that SMILES is non-directional, a single-directional recurrent neural network cannot precisely capture molecular information. Therefore, SGGRL utilizes Bi-LSTM units to preprocess the one-hot vectors and capture SMILES contextual information,
\begin{equation}
    \begin{cases}
    \overrightarrow{h_{s_i}}=\overrightarrow{\mathbf{LSTM}}(x_{s_i}, \overrightarrow{h_{s_{i-1}}})\\
    \overleftarrow{h_{s_i}}=\overleftarrow{\mathbf{LSTM}}(x_{s_i}, \overleftarrow{h_{s_{i-1}}}) 
    \end{cases}
    \label{eq:LSTM}
\end{equation}
where the $\overrightarrow{h_{s_i}}$ and $\overleftarrow{h_{s_i}}$ are the hidden states of token $s_i$, and the hidden state representation $h_{s_i}$ can be obtained by concatenating $\overrightarrow{h_{s_i}}$ and $\overleftarrow{h_{s_i}}$,
\begin{equation}
    h_{s_i} = \mathbf{CONCAT}(\overrightarrow{h_{s_i}}, \overleftarrow{h_{s_i}})
    \label{h_i generate}
\end{equation}

And then, the sequence encoding can be input to the transformer to learn the representation of SMILES. 

\textbf{Graph Encoder.} 
As shown in \cref{fig: Overview of SGGRL}~(a), a molecule can be described as a graph $G=(V, E)$, where $V$ is the set of atoms and $E$ is the set of bonds. To obtain the representation information from graphs, SGGRL introduces CMPNN~\cite{ijcai2020p392} as the basic graph encoder to obtain the representations of atoms. And the graph encoder of SGGRL consists of several CMPNN layers. The CMPNN module consists of the following two components: $\mathbf{AGGREGATE}$ and $\mathbf{COMMUNICATE}$. The input is the graph $G=(V, E)$, which includes node attributes $X_V$ and edge attributes $X_E$. And the initial node hidden representation $h^0_{v}=x_{v}$ and edge hidden representation $h^0_{e_{uv}}=x_{e_{uv}}$. They are propagated at $k$-th iteration as follows,
\begin{equation}
    m^k_v = \mathbf{AGGREGATE}(h^{k-1}_e)
\end{equation}
\begin{equation}
    h^k_v = \mathbf{COMMUNICATE}(m^k_v, h^{k-1}_v)
\end{equation}
\begin{equation}
    h^k_{e_{uv}} = \sigma(h^0_{e_{uv}} + W \cdot (h^k_v - h^{k-1}_{e_{uv}})))
\end{equation}
where $m^k_v$ is the message obtained by node $v$, $h^k_v$ and $h^k_{e_{uv}}$ is the hidden representation of them, $W$ is a learnable weight matrix, and $\sigma$ is the activation function. After $L$ iterations, once again the following operations are performed to obtain the final messages and node representations:
\begin{equation}
    m_v = \mathbf{AGGREGATE}(h^{L}_{e_{uv}})
\end{equation}
\begin{equation}
    h_v = \mathbf{COMMUNICATE}(m_v, h^L_v, x_v)
\end{equation}

The $\mathbf{AGGREGATE}(\cdot)$ contains a message booster that generates the maximum pooling result of the sum of the edge hidden representations $h_e$ and computes its elemental product with the sum of $h_e$. The $\mathbf{COMMUNICATE}(\cdot)$ takes the form of a multi-layer perception.

\textbf{Geometry Encoder.} 
Since combining geometric information into the graph structure can better predict molecular properties, SGGRL further introduces the geometry representation learning module. As is shown in \cref{fig: Overview of SGGRL}~(a), one can define a geometric graph $H=(V, E, A)$, where $v\in V$ is regarded as the node, $e\in E$ is the edge, and $a\in A$ is the bond angle. We obtain node representations of each input geometric graph by introducing GEMGNN~\cite{fang2022geometry}, which use  $\mathbf{AGGEGATE}$ and $\mathbf{COMBINE}$ functions of GIN~\cite{xu2018powerful}. 
And to better learn the bond information and fuse the atom and bond information of the geometric graphs, GINE~\cite{hu2020strategies} is introduced to replace GIN. Formally, the node representation $h^k_v$ can be expressed as:
\begin{equation}
    m^k_v = \mathbf{AGGEGATE}(h^{k-1}_v, h^{k-1}_u, h^{k-1}_{e_{uv}})
\end{equation}
\begin{equation}
    h^k_v = \mathbf{COMBINE}(h^{k-1}_v, m^k_v)
\end{equation}
where the $h^{k-1}_{e_{uv}}$ is the $(k-1)$-th layer edge representation. And the $k$-th layer edge representation can be expressed as:
\begin{equation}
\begin{aligned}
    m^k_{e_{uv}} &= \mathbf{AGGEGATE}(\\ 
    &\{h^{k-1}_{e_{uv}}, h^{k-1}_{e_{uw}},x_{a_{wuv}}:w\in\mathscr{N}(u)\}\bigcup\\ 
    &\{h^{k-1}_{e_{uv}}, h^{k-1}_{e_{vw}},x_{a_{uvw}}:w\in\mathscr{N}(v)\})
\end{aligned}
\end{equation}
\begin{equation}
    h^k_{e_{uv}} = \mathbf{COMBINE}(h^{k-1}_{e_{uv}}, m^k_{e_{uv}})
\end{equation}
where $w\in\mathscr{N}(u)$ and $w\in\mathscr{N}(v)$ refer to the neighbor of node $v$ and $u$ respectively. Note that the $h^0_v$ is encoded by the embedding layer with the features of node $v$, the $h^0_{e_{uv}}$ is encoded by the embedding layer with the features of edge $(u,v)$, and the $x_{a_{wuv}}$ is encoded by the embedding layer with the features of angle $a_{uvw}$.

Above all, one can get different modal representations of a token or an atom. Finally, to effectively capture the important information of different modalities and reduce the redundant information, SGGRL introduces a GlobalAttentionPool layer to acquire the molecular-level representation $H_s$, $H_g$, and $H_h$.




\subsection{Fusion Module}\label{Fusion}
To capture comprehensive information on different modalities, a fusion layer is designed to acquire a joint representation. And it can be expressed as:
\begin{equation}
    H_{fusion} = W_s\cdot H_s + W_g\cdot H_g + W_h\cdot H_h + b
\end{equation}
where $W_s$, $W_g$ and $W_h$ are learnable weights, and $b$ is a bias vector.
\subsection{Loss Function}\label{Loss}
Typically, the goal of representation learning is to evaluate the loss of supervised signals from downstream tasks, i.e., the molecular property labels in this work. However, since our model uses three different representation modules from sequences, graphs, and geometries, differences in the latent spaces of different models need to be eliminated. In order to maximize the consistency of feature space between different representations, SGGRL introduces a similarity-based contrastive mechanism. 
\subsubsection{Contrastive Loss}
In order to achieve the compatibility and consistency of different modalities, we introduce the NT-Xent~\cite{chen2020simple} contrastive loss for each pair of modalities. And the total contrastive loss of SGGRL can be expressed as:
\begin{equation}
\begin{aligned}
    \mathcal{L}_{cl}=&-\log{\frac{\mathrm{exp}(\mathrm{sim}(Z_{g(i)},Z_{h(i)})/T)}{\sum_{j=1,i\neq j}^{N}{\mathrm{exp}(\mathrm{sim}(Z_{g(i)}, Z_{h(j)})/T)}}} \\
    &-\log{\frac{\mathrm{exp}(\mathrm{sim}(Z_{s(i)},Z_{g(i)})/T)}{\sum_{j=1,i\neq j}^{N}{\mathrm{exp}(\mathrm{sim}(Z_{s(i)}, Z_{g(j)})/T)}}} \\
    &-\log{\frac{\mathrm{exp}(\mathrm{sim}(Z_{h(i)},Z_{s(i)})/T)}{\sum_{j=1,i\neq j}^{N}{\mathrm{exp}(\mathrm{sim}(Z_{h(i)}, Z_{s(j)})/T)}}}
\end{aligned}
\end{equation}
where $Z_s$, $Z_g$ and $Z_h$ is the output of a total projection head that maps the representations $H_s$, $H_g$ and $H_h$ to the space where contrastive loss is
applied, $\mathrm{sim}(m,n)= m^{\mathrm{T}}n/\begin{Vmatrix}m \\ \end{Vmatrix}\begin{Vmatrix}n \\ \end{Vmatrix}$ denotes a similarity calculation function, and $T$ is the contrastive loss rate.

\subsubsection{Joint Loss}
To predict molecular properties, we introduce a molecular property predictor $\mathcal{F}$ after the final fusion representation $H_{fusion}$. For different tasks, the loss function is different between predicted labels $y_{pred}$ and ground-truth labels $y$. Formally, they can be defined as follows:
\begin{equation}
    y_{pred} = \mathcal{F}(H_{fusion})
\end{equation}
\begin{equation}
    \mathcal{L}_{cls} = \mathcal{L}_{BCE}(y_{pred}, y) + \alpha \mathcal{L}_{cl}
\end{equation}
\begin{equation}
    \mathcal{L}_{reg} = \mathcal{L}_{MSE}(y_{pred}, y) + \alpha \mathcal{L}_{cl}
\end{equation}
where the $\mathcal{F}$ consists of $n$ layers multi-layer perception, $\mathcal{L}_{BCE}$ is the BCEWithLogitsLoss and $\mathcal{L}_{MSE}$ is the MSELoss. Due to the different magnitudes of contrastive loss and prediction task loss, we introduce the ratio $\alpha$ of the contrastive loss to balance different losses and make the training more stability

\section{Experiments}
In this section, lots of experiments are conducted to demonstrate the performance of SGGRL on seven benchmark datasets. Specifically, we describe the datasets, baselines, and settings used in the experiments, give overall results, and present the ablation experiments. Finally, we conduct the data distribution visualization experiments learned by different models for more intuitive comparison.

\subsection{Datasets}
In this section, we utilize seven benchmark datasets from MoleculeNet~\cite{wu2018moleculenet} to conduct experiments on classification and regression tasks. \cref{tab: dataset} shows the statistics of these datasets. Among them, BACE and BBBP are used for single-classification tasks, SIDER, ClinTox, and Tox21 are used for multi-task classification, and FreeSolv and ESOL are used for regression tasks.

\begin{table}[!t]
    \caption{Summary of datasets.}
    \centering
    \renewcommand{\arraystretch}{1.3}
    \resizebox{1\columnwidth}{!}{
    \begin{tabular}{cccccc}
    \toprule 
    \textbf{Dataset} & \textbf{\# Graphs} & \textbf{Avg. \# Nodes} & \textbf{Avg. \# Edges} & \textbf{\# Tasks} & \textbf{\# Task Type}\\
    \hline 
    BACE & 1513 & 34.1 & 36.9 & 1 & Classification \\
    BBBP & 2039 & 24.1 & 26.0 & 1 & Classification \\
    SIDER & 1427 & 33.6 & 35.4 & 27 & Classification \\
    ClinTox & 1478 & 26.2 & 27.9 & 2 & Classification \\
    Tox21 & 7831 & 18.6 & 19.3 & 12 & Classification \\
    FreeSolv & 642 & 8.7 & 8.4 & 1 & Regression \\
    ESOL & 1128 & 13.3 & 13.7 & 1 & Regression \\
    \bottomrule 
    \end{tabular}
    }
    \label{tab: dataset}
\end{table}

\begin{table}[!t]
    \caption{Summary of the hyper-parameter ranges used in SGGRL.}
    \centering
    \renewcommand{\arraystretch}{1.3}
    \resizebox{1\columnwidth}{!}{
    \begin{tabular}{ccc}
    \toprule 
    \textbf{Hyper-Parameters} & \textbf{ Descriptions} & \textbf{ Range} \\
    \hline 
    $\alpha$ & contrastive loss ratio & 0.1 \\
    Epoch & number of epochs for training & 100 \\
    Batch size & batch size used during the training of the model & 64, 128, 256 \\
    Max learning rate & max learning rate of Noam LR scheduler & 2e-3, 2e-4 \\ 
    Final learning rate & final learning rate of Noam LR scheduler & 1e-3, 1e-4 \\
    Initial learning rate & initial learning rate of Noam LR scheduler & 1e-3, 1e-4 \\
    \hline
    Layers (CMPNN) & the number of CMPNN layers & 5 \\
    Layers (GEMGNN) & the number of GemGNN layers & 8 \\
    Layers (Bi-LSTM) & the number of Bi-LSTM layers & 3 \\
    Layers (Transformer) & the number of Transformer layers & 4 \\
    Hidden dim (CMPNN) & the hidden dimension of CMPNN & 256 \\
    Hidden dim (GemGNN) & the hidden dimension of GemGNN & 256 \\
    Hidden dim (Bi-LSTM) & the hidden dimension of Bi-LSTM & 256 \\
    Hidden dim (Transformer) & the hidden dimension of Transformer & 256 \\
    Head number (Transformer) & the number of self-attention heads in Transformer & 4 \\
    \bottomrule 
    \end{tabular}
    }
    \label{tab: hyperparameters}
\end{table}
\subsection{Baselines}
We compare our SGGRL method with nine baselines, including sequence-based, graph-based, geometry-based, and multi-modal methods. Specifically, all of the baseline methods are summarized as follows:
\begin{itemize}
\item RNNS2S~\cite{xu2017seq2seq}: An unsupervised molecular representation method that provides a continuous feature vector for each molecule based on Gated Recurrent Unit (GRU). 
\item ST (SMILES Transformer)~\cite{honda2019smiles}: A data-driven molecular fingerprinting method that learns molecular fingerprints by unsupervised pretraining of sequence language models based on Transformer.
\item GIN~\cite{xu2018powerful}: A graph representation learning framework based on graph isomorphism networks.
\item CMPNN~\cite{song2020communicative}: A communication message passing neural network that enhances message interaction between atoms and bonds through a communicative kernel.
\item GEM~\cite{fang2022geometry}: A geometrically enhanced molecular representation-based learning approach.
\item Uni-Mol~\cite{zhou2023unimol}: A 3D molecular representation learning framework that can predict 3D positions.
\item GraSeq~\cite{guo2020graseq}: A fusion representation learning model for graphs and sequences that utilizes joint information to learn efficient molecular representations for different downstream tasks in molecular property prediction.
\item 3D Infomax~\cite{stark20223d}: A pre-training method for molecular property prediction that maximizes the mutual information between learned 3D vectors and graph neural network representations. 
\item GraphMVP\cite{liu2021pre}: A Graph Multi-View Pre-training method that learns molecular graph representations by utilizing correspondence and consistency between 2D topological structures and 3D geometric views.
\end{itemize}

\subsection{Experimental Settings}
For all experiments, the datasets are split by random splitting and scaffold splitting according to MoleculeNet~\cite{wu2018moleculenet}. The ratio of train, validation, and test sets for these two splitting ways is 8:1:1. For a fair comparison, we perform five independent runs with different random seeds, and calculate the mean and standard deviation of (ROC-AUC) and root-mean-squared error (RMSE) metrics. The hyper-parameters used in the experiments are demonstrated in \cref{tab: hyperparameters}.

\begin{table*}[!t]
  \centering
  \caption{Molecular property prediction results on several classification and regression datasets with baselines and SGGRL.}
  \renewcommand{\arraystretch}{1.3}
  \resizebox{2\columnwidth}{!}{
    \begin{tabular}{c|ccccc|cc}
    \bottomrule
    \multirow{2}{*}{Methods} & \multicolumn{5}{c|}{ROC-AUC$\uparrow$ (Higher is better)} & \multicolumn{2}{c}{RMSE$\downarrow$ (Lower is better)} \\
    \cline{2-8}
    & BACE & BBBP & ClinTox & Sider & Tox21 & ESOL & FreeSolv \\
    \hline
    RNNS2S & 0.740$\pm$0.017 & 0.898$\pm$0.011 & 0.910$\pm$0.036 & 0.550$\pm$0.006 & 0.702$\pm$0.005 & 1.277$\pm$0.066 & 2.939$\pm$0.175  \\
    ST   & 0.727$\pm$0.026 & 0.910$\pm$0.010 & \underline{0.930$\pm$0.037} & 0.558$\pm$0.005 & 0.708$\pm$0.005 & 1.056$\pm$0.054 & 2.281$\pm$0.236  \\
    \hline
    GIN  & 0.795$\pm$0.034 & 0.792$\pm$0.059 & 0.694$\pm$0.024 & 0.591$\pm$0.016 & 0.806$\pm$0.002 & 0.885$\pm$0.051 & 1.619$\pm$0.202  \\
    CMPNN & \underline{0.873$\pm$0.029} & 0.927$\pm$0.017 & 0.901$\pm$0.016 & 0.639$\pm$0.041 & \underline{0.837$\pm$0.009} & 0.798$\pm$0.112 & \underline{1.570$\pm$0.442}  \\
    \hline
    GEM  & 0.856$\pm$0.011 & 0.724$\pm$0.004 & 0.901$\pm$0.013 & \underline{0.672$\pm$0.004} & 0.781$\pm$0.001 & 0.798$\pm$0.029 & 1.877$\pm$0.094  \\
    Uni-Mol & 0.857$\pm$0.020 & 0.729$\pm$0.060 & 0.919$\pm$0.180 & 0.659$\pm$0.130 & 0.796$\pm$0.050 & \underline{0.788$\pm$0.029} & 1.620$\pm$0.035  \\
    \hline
    GraSeq & 0.764$\pm$0.002 & \underline{0.932$\pm$0.015} & 0.606$\pm$0.030 & 0.578$\pm$0.024 & 0.802$\pm$0.005 & 1.258$\pm$0.004 & 2.746$\pm$0.012  \\
    3D Infomax & 0.794$\pm$0.019 & 0.691$\pm$0.010 & 0.594$\pm$0.032 & 0.534$\pm$0.033 & 0.745$\pm$0.074 & 0.894$\pm$0.028 & 2.337$\pm$0.227  \\
    GraphMVP & 0.812$\pm$0.090 & 0.724$\pm$0.016 & 0.775$\pm$0.042 & 0.639$\pm$0.012 & 0.744$\pm$0.020 & 1.029$\pm$0.033 & 1.893$\pm$0.063  \\
    \hline
    \textbf{SGGRL(Ours)} & \textbf{0.917$\pm$0.020} & \textbf{0.967$\pm$0.010}  & \textbf{0.979$\pm$0.016} &  \textbf{0.682$\pm$0.015}  & \textbf{0.847$\pm$0.013}  & \textbf{0.628$\pm$0.057} &  \textbf{0.847$\pm$0.116} \\
    \toprule
    \end{tabular}%
    }
  \label{tab: result}%
\end{table*}%

\begin{table*}[!t]
  \centering
  \caption{Ablation experiments results on seven benchmark datasets}
  \renewcommand{\arraystretch}{1.3}
  \resizebox{2\columnwidth}{!}{
    \begin{tabular}{ccccc|ccccccc}
    \bottomrule
    Sequence & Graph & Geometry & Bi-LSTM & AttentionPool & BACE  & BBBP  & Clintox & Sider & Tox21 & ESOL  & Freesolv \\
    \hline
    \textbf{-}   & \CheckmarkBold & \CheckmarkBold & \CheckmarkBold & \CheckmarkBold & 0.818$\pm$0.058 & 0.945$\pm$0.003 & 0.900$\pm$0.091 & 0.636$\pm$0.008  &  0.755$\pm$0.016 & 0.790$\pm$0.112  & 0.870$\pm$0.134 \\
    \CheckmarkBold & \textbf{-}  & \CheckmarkBold & \CheckmarkBold & \CheckmarkBold &  0.866$\pm$0.032   & 0.948$\pm$0.002  & 0.941$\pm$0.099 & 0.628$\pm$0.020 & 0.791$\pm$0.023 & 0.802$\pm$0.105 & 1.271$\pm$0.523 \\
    \CheckmarkBold & \CheckmarkBold & \textbf{-}  & \CheckmarkBold & \CheckmarkBold & 0.876$\pm$0.018 & 0.914$\pm$0.036  & 0.932$\pm$0.020 & 0.636$\pm$0.009 & 0.777$\pm$0.021 & 0.839$\pm$0.074 & 0.927$\pm$0.020 \\
    \CheckmarkBold & \CheckmarkBold & \CheckmarkBold & \textbf{-}  & \CheckmarkBold & 0.847$\pm$0.053  & 0.942$\pm$0.009 & 0.932$\pm$0.063  & 0.655$\pm$0.036 & 0.822$\pm$0.008  & 0.772$\pm$0.118 & 1.535$\pm$0.049 \\
    \CheckmarkBold & \CheckmarkBold & \CheckmarkBold & \CheckmarkBold & \textbf{-}  & 0.893$\pm$0.008 & 0.904$\pm$0.028 & 0.920$\pm$0.011 & 0.620$\pm$0.028 & 0.780$\pm$0.031 & 0.738$\pm$0.153 & 0.864$\pm$0.138 \\
    \CheckmarkBold & \CheckmarkBold & \CheckmarkBold & \CheckmarkBold & \CheckmarkBold & \textbf{0.917$\pm$0.020} & \textbf{0.967$\pm$0.010}  & \textbf{0.979$\pm$0.016} &  \textbf{0.682$\pm$0.015}  & \textbf{0.847$\pm$0.013}  & \textbf{0.628$\pm$0.057} &  \textbf{0.847$\pm$0.116}\\
    \toprule
    \end{tabular}%
    }
  \label{tab: ablation}%
\end{table*}%

\subsection{Results and Analysis}
Comparison experiments were conducted on five classification datasets and two regression datasets to demonstrate the effectiveness of SGGRL. The results of SGGRL and nine baselines are reported in Table \cref{tab: result}, where the best-performing results are highlighted in bold and the suboptimal results are underlined. It is easy to conclude that:
1) By combining different modalities of molecules, SGGRL significantly outperforms the baselines on all benchmark datasets. Specifically, SGGRL achieves $91.7\%$ accuracy on BACE, $96.7\%$ on BBBP, $97.9\%$ on Clintox, $68.2\%$ on Sider, and $84.7\%$ on Tox21, with an average relative improvement of $3.35\%$ compared to suboptimal models in the classification tasks. Moreover, its regression mean squared error achieves an average decrease of $0.441$ compared to suboptimal models. 2) SSGGRL also achieves state-of-the-art performance on all benchmark datasets. In comparison to GraSeq, SGGRL introduces the transformer module and the geometry modality, both of which effectively enhance the representation with $32.6\%$ relative improvement. Additionally, SGGRL introduces the consistency learning module and achieves more effective modalities fusion. Compared with 3D Infomax and GraphMVP, SGGRL introduces the sequence modality to capture the chirality and functional group information, enhancing the representation learning and achieving $31.6\%$ relative improvement. 3) Whether for single-task or multi-task datasets, SGGRL consistently achieves excellent performance, especially on small-scale datasets such as BACE and FreeSolv. Furthermore, the stability of SGGRL outperforms the baselines in most cases. These phenomena further demonstrate that multi-modal molecular representation learning can more comprehensively capture molecular characteristics and enhance molecular representation.

\subsection{Ablation Studies}
Here, lots of ablation experiments were conducted, which aim to further verify the effectiveness of the multi-modal mechanism, GlobalAttentionPool Readout layer, and Bi-LSTM preprocessed encoding operation. Table \cref{tab: ablation} reports the mean and standard deviation of ROC-AUC and RMSE values. Among them, the \textbf{-} in sequence, graph, and geometry columns refers to the ablation methods that remove the corresponding modal; 
the \textbf{-} in the Bi-LSTM column refers to the ablation method that uses the embedding layer and position embedding layer to preprocess sequence features before the transformer; 
the \textbf{-} in the GlobalAttentionPool column refers to the ablation method that uses GlobalMeanPool in the molecular encoder readout layer. 
Based on the observations in Table~\cref{tab: ablation}, it is easy to conclude that: 1) SGGRL significantly outperforms the bimodal methods, which effectively verifies that the utilization of sequence, graph, and geometry modal can more comprehensively capture the molecular features and information. On the whole, the contribution of the geometry modality to the SGGRL is small, which may be because the molecular conformation is randomized.
2) SGGRL method is more powerful than the embedding layer and position embedding layer-based method. It makes sense that the smiles sequence is undirect and the Bi-LSTM can extract bidirectional information. 
3) SGGRL is superior to the GlobalMeanPool based. It can be contributed that the AttentionPool layer pays attention to the more important information of different modalities, which releases the redundancy problem. Above all, these observations re-confirm the effectiveness of SGGRL.

\begin{figure*}[!t]
    \centering
    \includegraphics[width=\linewidth]{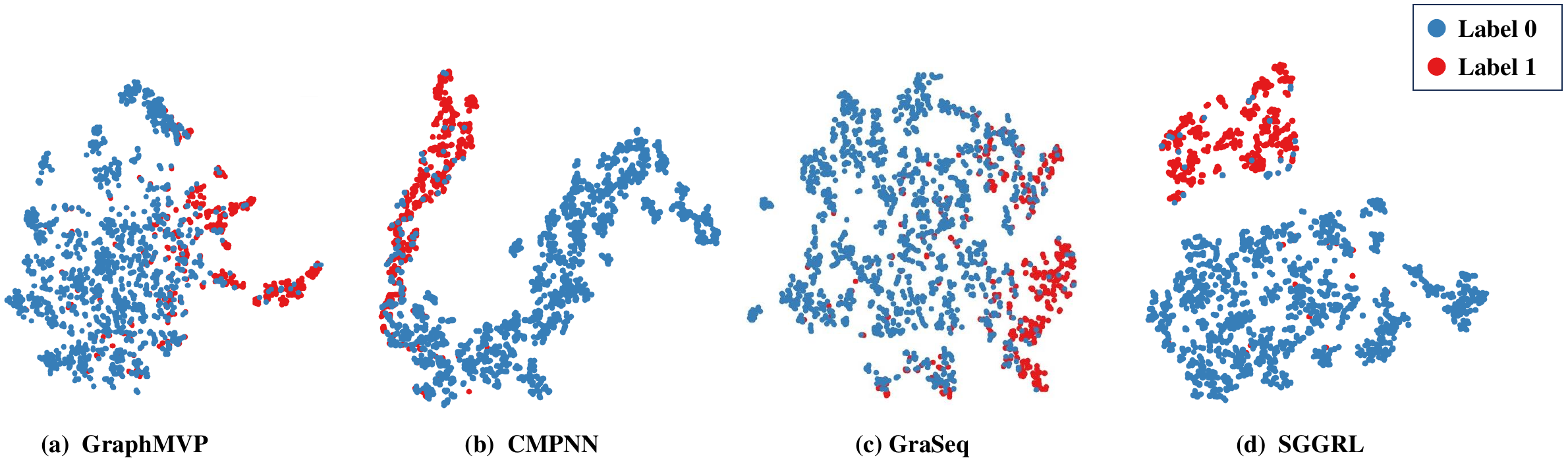}
    \caption{T-SNE visualization of the molecular representation space of GraphMVP, CMPNN, GraSeq, and SGGRL on the BBBP dataset. The red dots denote the negative labels, and the blue dots denote the positive labels.}
    \label{fig: Molecular Representation Visualization}
\end{figure*}

\subsection{Molecular Representation Visualization}
To intuitively showcase the representative capability of SGGRL, we visualize the molecular representations learned by GraphMVP, CMPNN, GraSeq, and SGGRL on the BBBP dataset using T-SNE~\cite{van2008visualizing}. Based on the observations in \cref{fig: Molecular Representation Visualization}, it is evident that four models achieve a remarkable separation of molecular properties, with CMPNN, GraSeq, and SGGRL producing clearer results. 
Notably, GraphMVP solely utilizes geometry information during pre-training, lacking the fusion mechanism needed to capture multi-modal information. Additionally, the inclusion of randomized geometry information may introduce noise, resulting in adverse effects, which also provides verification of the analysis in the ablation study. 
The edge-to-node message-passing mechanism enables CMPNN to acquire comprehensive molecular semantics; however, its performance is limited due to single-modality learning. 
Furthermore, the classification boundary of SGGRL is the most distinct. This reaffirms that SGGRL combined information from different modalities can lead to a more accurate capture of molecular semantics. As SGGRL was purposefully designed to incorporate sequence, graph, and geometry modalities, it yields comprehensive molecular representations and achieves the best performance.

\section{Conclusion}
In this paper, we propose SGGRL, a novel molecular representation learning model based on the multi-modals of molecules for molecular property prediction. Specifically, SGGRL is comprised of an encoder module, readout module, self-supervised learning module, and property prediction module. Among them, the encoder module utilizes sequence-based, graph-based, and geometry-based encoders to acquire molecular representations which contain different modal information. The readout layer consists of a GlobalAttentionPool Readout layer which aims at capturing important information of different modals to acquire molecule-level representations and a weighted multi-modal fusion layer which aims at fusing different modal representations with attention mechanisms to get the final molecular representation. Moreover, to achieve the compatibility and consistency of different modal representations, SGGRL introduced a multi-modal contrastive learning task module. Experiments demonstrate that SGGRL achieves state-of-the-art performance compared with several baselines in molecular property prediction tasks, and shows its competitiveness. 

In the future, we will continue to explore the molecular representation works and improve the SGGRL. Specifically, the modal encoders and the fusion mechanisms are two crucial modules of the multi-modal molecular representation model. The more representative encoders and more effective fusion mechanisms can help the multi-modal model achieve a more accurate molecular representation. In this work, SGGRL considers the independent modal encoding process with the late fusion of different modals and achieves strong portability. More encoding interaction in the encoding process can be taken into consideration to capture more comprehensive multi-modal information. And more applications of SGGRL for other biochemical tasks will be explored.

\bibliographystyle{IEEEtran}
\bibliography{ref}

\begin{thebibliography}{10}
\providecommand{\url}[1]{#1}
\csname url@samestyle\endcsname
\providecommand{\newblock}{\relax}
\providecommand{\bibinfo}[2]{#2}
\providecommand{\BIBentrySTDinterwordspacing}{\spaceskip=0pt\relax}
\providecommand{\BIBentryALTinterwordstretchfactor}{4}
\providecommand{\BIBentryALTinterwordspacing}{\spaceskip=\fontdimen2\font plus
\BIBentryALTinterwordstretchfactor\fontdimen3\font minus \fontdimen4\font\relax}
\providecommand{\BIBforeignlanguage}[2]{{%
\expandafter\ifx\csname l@#1\endcsname\relax
\typeout{** WARNING: IEEEtran.bst: No hyphenation pattern has been}%
\typeout{** loaded for the language `#1'. Using the pattern for}%
\typeout{** the default language instead.}%
\else
\language=\csname l@#1\endcsname
\fi
#2}}
\providecommand{\BIBdecl}{\relax}
\BIBdecl

\bibitem{schneider2020rethinking}
P.~Schneider, W.~P. Walters, A.~T. Plowright, N.~Sieroka, J.~Listgarten, R.~A. Goodnow~Jr, J.~Fisher, J.~M. Jansen, J.~S. Duca, T.~S. Rush \emph{et~al.}, ``Rethinking drug design in the artificial intelligence era,'' \emph{Nature Reviews Drug Discovery}, vol.~19, no.~5, pp. 353--364, 2020.

\bibitem{li2022deep}
Z.~Li, M.~Jiang, S.~Wang, and S.~Zhang, ``Deep learning methods for molecular representation and property prediction,'' \emph{Drug Discovery Today}, p. 103373, 2022.

\bibitem{walters2020applications}
W.~P. Walters and R.~Barzilay, ``Applications of deep learning in molecule generation and molecular property prediction,'' \emph{Accounts of chemical research}, vol.~54, no.~2, pp. 263--270, 2020.

\bibitem{fang2022geometry}
X.~Fang, L.~Liu, J.~Lei, D.~He, S.~Zhang, J.~Zhou, F.~Wang, H.~Wu, and H.~Wang, ``Geometry-enhanced molecular representation learning for property prediction,'' \emph{Nature Machine Intelligence}, vol.~4, no.~2, pp. 127--134, 2022.

\bibitem{zhang2018seq3seq}
X.~Zhang, S.~Wang, F.~Zhu, Z.~Xu, Y.~Wang, and J.~Huang, ``Seq3seq fingerprint: towards end-to-end semi-supervised deep drug discovery,'' in \emph{Proceedings of the 2018 ACM international conference on bioinformatics, computational biology, and health informatics}, 2018, pp. 404--413.

\bibitem{wieder2020compact}
O.~Wieder, S.~Kohlbacher, M.~Kuenemann, A.~Garon, P.~Ducrot, T.~Seidel, and T.~Langer, ``A compact review of molecular property prediction with graph neural networks,'' \emph{Drug Discovery Today: Technologies}, vol.~37, pp. 1--12, 2020.

\bibitem{weininger1988smiles}
D.~Weininger, ``Smiles, a chemical language and information system. 1. introduction to methodology and encoding rules,'' \emph{Journal of chemical information and computer sciences}, vol.~28, no.~1, pp. 31--36, 1988.

\bibitem{xu2017seq2seq}
Z.~Xu, S.~Wang, F.~Zhu, and J.~Huang, ``Seq2seq fingerprint: An unsupervised deep molecular embedding for drug discovery,'' in \emph{Proceedings of the 8th ACM international conference on bioinformatics, computational biology, and health informatics}, 2017, pp. 285--294.

\bibitem{wang2019smiles}
S.~Wang, Y.~Guo, Y.~Wang, H.~Sun, and J.~Huang, ``Smiles-bert: large scale unsupervised pre-training for molecular property prediction,'' in \emph{Proceedings of the 10th ACM international conference on bioinformatics, computational biology and health informatics}, 2019, pp. 429--436.

\bibitem{honda2019smiles}
S.~Honda, S.~Shi, and H.~R. Ueda, ``Smiles transformer: Pre-trained molecular fingerprint for low data drug discovery,'' \emph{arXiv preprint arXiv:1911.04738}, 2019.

\bibitem{ijcai2020p392}
Y.~Song, S.~Zheng, Z.~Niu, Z.-h. Fu, Y.~Lu, and Y.~Yang, ``Communicative representation learning on attributed molecular graphs,'' in \emph{Proceedings of the Twenty-Ninth International Joint Conference on Artificial Intelligence}, C.~Bessiere, Ed.\hskip 1em plus 0.5em minus 0.4em\relax International Joint Conferences on Artificial Intelligence Organization, 7 2020, pp. 2831--2838, main track.

\bibitem{gilmer2017neural}
J.~Gilmer, S.~S. Schoenholz, P.~F. Riley, O.~Vinyals, and G.~E. Dahl, ``Neural message passing for quantum chemistry,'' in \emph{International conference on machine learning}.\hskip 1em plus 0.5em minus 0.4em\relax PMLR, 2017, pp. 1263--1272.

\bibitem{gasteiger2021gemnet}
J.~Gasteiger, F.~Becker, and S.~G{\"u}nnemann, ``Gemnet: Universal directional graph neural networks for molecules,'' \emph{Advances in Neural Information Processing Systems}, vol.~34, pp. 6790--6802, 2021.

\bibitem{Shi2020GraphAF}
C.~Shi*, M.~Xu*, Z.~Zhu, W.~Zhang, M.~Zhang, and J.~Tang, ``Graphaf: a flow-based autoregressive model for molecular graph generation,'' in \emph{International Conference on Learning Representations}, 2020.

\bibitem{hao2020asgn}
Z.~Hao, C.~Lu, Z.~Huang, H.~Wang, Z.~Hu, Q.~Liu, E.~Chen, and C.~Lee, ``Asgn: An active semi-supervised graph neural network for molecular property prediction,'' in \emph{Proceedings of the 26th ACM SIGKDD International Conference on Knowledge Discovery \& Data Mining}, 2020, pp. 731--752.

\bibitem{stark20223d}
H.~St{\"a}rk, D.~Beaini, G.~Corso, P.~Tossou, C.~Dallago, S.~G{\"u}nnemann, and P.~Li{\`o}, ``3d infomax improves gnns for molecular property prediction,'' in \emph{International Conference on Machine Learning}.\hskip 1em plus 0.5em minus 0.4em\relax PMLR, 2022, pp. 20\,479--20\,502.

\bibitem{wang2023automated}
X.~Wang, H.~Zhao, W.-w. Tu, and Q.~Yao, ``Automated 3d pre-training for molecular property prediction,'' in \emph{Proceedings of the 29th ACM SIGKDD Conference on Knowledge Discovery and Data Mining}, 2023, pp. 2419--2430.

\bibitem{guo2020graseq}
Z.~Guo, W.~Yu, C.~Zhang, M.~Jiang, and N.~V. Chawla, ``Graseq: graph and sequence fusion learning for molecular property prediction,'' in \emph{Proceedings of the 29th ACM international conference on information \& knowledge management}, 2020, pp. 435--443.

\bibitem{li2022geomgcl}
S.~Li, J.~Zhou, T.~Xu, D.~Dou, and H.~Xiong, ``Geomgcl: Geometric graph contrastive learning for molecular property prediction,'' in \emph{Proceedings of the AAAI conference on artificial intelligence}, vol.~36, no.~4, 2022, pp. 4541--4549.

\bibitem{cereto2015molecular}
A.~Cereto-Massagu{\'e}, M.~J. Ojeda, C.~Valls, M.~Mulero, S.~Garcia-Vallv{\'e}, and G.~Pujadas, ``Molecular fingerprint similarity search in virtual screening,'' \emph{Methods}, vol.~71, pp. 58--63, 2015.

\bibitem{rupp2012fast}
M.~Rupp, A.~Tkatchenko, K.-R. M{\"u}ller, and O.~A. Von~Lilienfeld, ``Fast and accurate modeling of molecular atomization energies with machine learning,'' \emph{Physical review letters}, vol. 108, no.~5, p. 058301, 2012.

\bibitem{zheng2019identifying}
S.~Zheng, X.~Yan, Y.~Yang, and J.~Xu, ``Identifying structure--property relationships through smiles syntax analysis with self-attention mechanism,'' \emph{Journal of chemical information and modeling}, vol.~59, no.~2, pp. 914--923, 2019.

\bibitem{DMPNN2019}
K.~Yang, K.~Swanson, W.~Jin, C.~Coley, P.~Eiden, H.~Gao, A.~Guzman-Perez, T.~Hopper, B.~Kelley, M.~Mathea \emph{et~al.}, ``Analyzing learned molecular representations for property prediction,'' \emph{Journal of chemical information and modeling}, vol.~59, no.~8, pp. 3370--3388, 2019.

\bibitem{axelrod2022geom}
S.~Axelrod and R.~Gomez-Bombarelli, ``Geom, energy-annotated molecular conformations for property prediction and molecular generation,'' \emph{Scientific Data}, vol.~9, no.~1, p. 185, 2022.

\bibitem{mao2021molecular}
K.~Mao, X.~Xiao, T.~Xu, Y.~Rong, J.~Huang, and P.~Zhao, ``Molecular graph enhanced transformer for retrosynthesis prediction,'' \emph{Neurocomputing}, vol. 457, pp. 193--202, 2021.

\bibitem{liu2022pretraining}
S.~Liu, H.~Wang, W.~Liu, J.~Lasenby, H.~Guo, and J.~Tang, ``Pre-training molecular graph representation with 3d geometry,'' in \emph{International Conference on Learning Representations}, 2022.

\bibitem{xu2018powerful}
K.~Xu, W.~Hu, J.~Leskovec, and S.~Jegelka, ``How powerful are graph neural networks?'' \emph{arXiv preprint arXiv:1810.00826}, 2018.

\bibitem{hu2020strategies}
W.~Hu, B.~Liu, J.~Gomes, M.~Zitnik, P.~Liang, V.~Pande, and J.~Leskovec, ``Strategies for pre-training graph neural networks,'' in \emph{International Conference on Learning Representations (ICLR)}, 2020.

\bibitem{chen2020simple}
T.~Chen, S.~Kornblith, M.~Norouzi, and G.~Hinton, ``A simple framework for contrastive learning of visual representations,'' in \emph{International conference on machine learning}.\hskip 1em plus 0.5em minus 0.4em\relax PMLR, 2020, pp. 1597--1607.

\bibitem{wu2018moleculenet}
Z.~Wu, B.~Ramsundar, E.~N. Feinberg, J.~Gomes, C.~Geniesse, A.~S. Pappu, K.~Leswing, and V.~Pande, ``Moleculenet: a benchmark for molecular machine learning,'' \emph{Chemical science}, vol.~9, no.~2, pp. 513--530, 2018.

\bibitem{song2020communicative}
Y.~Song, S.~Zheng, Z.~Niu, Z.-H. Fu, Y.~Lu, and Y.~Yang, ``Communicative representation learning on attributed molecular graphs.'' in \emph{IJCAI}, vol. 2020, 2020, pp. 2831--2838.

\bibitem{zhou2023unimol}
G.~Zhou, Z.~Gao, Q.~Ding, H.~Zheng, H.~Xu, Z.~Wei, L.~Zhang, and G.~Ke, ``Uni-mol: A universal 3d molecular representation learning framework,'' in \emph{The Eleventh International Conference on Learning Representations}, 2023.

\bibitem{liu2021pre}
S.~Liu, H.~Wang, W.~Liu, J.~Lasenby, H.~Guo, and J.~Tang, ``Pre-training molecular graph representation with 3d geometry,'' \emph{arXiv preprint arXiv:2110.07728}, 2021.

\bibitem{van2008visualizing}
L.~Van~der Maaten and G.~Hinton, ``Visualizing data using t-sne.'' \emph{Journal of machine learning research}, vol.~9, no.~11, 2008.

\end{thebibliography}

\end{document}